\documentclass{aa}
\usepackage[varg]{txfonts}
\usepackage{graphicx}
\bibpunct{(}{)}{;}{a}{}{,} 

\begin{document}
\title{A Keplerian disk around a Class 0 source: ALMA observations of VLA1623A}

\author{Nadia M. Murillo\inst{1} 
\and Shih-Ping Lai\inst{2, 3} 
\and Simon Bruderer\inst{1} 
\and Daniel Harsono\inst{4, 5} 
\and Ewine F. van Dishoeck\inst{1,4}}

\institute{Max-Planck-Insitut f\"{u}r extraterrestrische Physik, Geissenbachstra\ss e 1, 85748, Garching bei Munchen, Germany
\and Institute of Astronomy and Department of Physics, National Tsing Hua University, 101 Section 2 Kuang Fu Road, Hsinchu 30013, Taiwan
\and Academia Sinica Institute of Astronomy and Astrophysics, P.O. Box 23-141, Taipei 10617, Taiwan
\and Leiden Observatory, Leiden University, Niels Bohrweg 2, 2300 RA, Leiden, the Netherlands
\and SRON Netherlands Institute for Space Research, P.O. Box 800, 9700 AV Groningen, The Netherlands}

\date{}

\abstract{Rotationally supported disks are critical in the star formation process. The questions of when do they form and what factors influence or hinder their formation have been studied but are largely unanswered. Observations of early stage YSOs are needed to probe disk formation.}{VLA1623 is a triple non-coeval protostellar system, with a weak magnetic field perpendicular to the outflow, whose Class 0 component, VLA1623A, shows a disk-like structure in continuum with signatures of rotation in line emission. We aim to determine whether this structure is in part or in whole a rotationally supported disk, i.e. a Keplerian disk, and what are its characteristics.}{ALMA Cycle 0 Early Science 1.3 mm continuum and C$^{18}$O (2-1) observations in the extended configuration are presented here and used to perform an analysis of the disk-like structure using PV diagrams and thin disk modelling with the addition of foreground absorption.}{The PV diagrams of the C$^{18}$O line emission suggest the presence of a rotationally supported component with a radius of at least 50 AU. Kinematical modelling of the line emission shows that the disk out to 180 AU is actually rotationally supported, with the rotation being well described by Keplerian rotation out to at leat 150 AU, and the central source mass to be $\sim$0.2 M$_{\sun}$ for an inclination of 55$^{\circ}$. Pure infall and conserved angular momentum rotation models are excluded.}{VLA1623A, a very young Class 0 source, presents a disk with an outer radius $R_{\rm out}$ = 180 AU with a Keplerian velocity structure out to at least 150 AU. The weak magnetic fields and recent fragmentation in this region of $\rho$ Ophiuchus may have played a lead role in the formation of the disk.}


\titlerunning{A Keplerian disk around a Class 0 source: ALMA observations of VLA1623A}
\authorrunning{N. M. Murillo et al.}

\maketitle

\section{Introduction}
\label{secintro}
Disks are key actors in the formation of stars. They are crucial for accretion and angular momentum distribution in the early stages and planet formation in the later stages. Rotationally supported disks have been observed using molecular lines in the Class II stage of star formation \citep{mannings1997, guilloteau1998, guilloteau1999, qi2004, hughes2009, rodriguez2010} while continuum disk-like structures, the so-called pseudo-disks, are reported in Class 0 objects \citep{jorgensen2009, enoch2009, enoch2011}. This leads to the expectation that rotationally supported or Keplerian disks evolve from pseudo-disks between the Class 0 and II stages. 

\begin{table*}
\caption{VLA1623's 1.3 mm continuum fluxes measured with ALMA}
\label{tabcont}
\centering
\begin{tabular}{c c c c c}
\hline \hline
Source & R.A. & Decl. & Peak (mJy beam$^{-1}$) & Integrated (mJy) \\
\hline
VLA1623A & 16:26:26.390 & -24:24:30.688 & 93.1 & 201.3$\pm$1.2 \\
VLA1623B & 16:26:26.309 & -24:24:30.588 & 93.5 & 96.3$\pm$0.7 \\
VLA1623W & 16:26:25.636 & -24:24:29.488 & 21.6 & 37.3$\pm$0.9 \\
\hline
\end{tabular}
\end{table*}

Idealized, non-magnetized conditions for a collapsing isothermal core suggest that a small rotating disk ($R_{\rm out} < 100$ AU) should form as early as the Class 0 stage and grow as $R \propto t^{3}$, where $t$ is time since collapse \citep{terebey1984}. Addition of magnetic fields to the problem offers varied results. Ideal MHD shows that magnetic field breaking can hinder disk formation \citep{mellon2008}, however if the magnetic field and rotation axes are misaligned, rotationally supported disks may form since the magnetic breaking efficiency is reduced \citep{hennebelle2009,krumholz2013}. Consideration of non-ideal MHD effects and their role in disk formation has been explored but there is no clear solution \citep{li2011}. 

Despite the observational relations and predictions obtained from simulations, it is still unclear when disks actually begin to form, and to what degree do factors such as magnetic fields and fragmentation hinder or encourage the formation of rotationally supported disks.

While observations of (sub-) Keplerian disks in Class I YSOs \citep{hogerheijde2001, brinch2007, lommen2008, jorgensen2009, takakuwa2012, yen2013} support the expectation that disks form between the Class 0 and II stages, recent observations have found indications that rotationally supported disks may be present as early as the Class 0 stage (NGC1333 IRAS4A: \citealt{choi2010}; L1527: \citealt{tobin2012}; VLA1623: \citealt{murillo2013}). Interestingly NGC1333 IRAS4A \citep{girart2006} and L1527 \citep{davidson2011} present hour-glass-like magnetic fields, while VLA1623 only shows a large scale "weak" magnetic field perpendicular to the large-scale outflow direction \citep{holland1996, hull2013}. Additionally, while NGC1333 IRAS4A and VLA1623 are confirmed multiples, it is unclear whether L1527 is a protobinary \citep{loinard2002} or a single protostar. This brings back the question of what factors play a role in the formation of a Keplerian disk and if the combination of these factors is more relevant than the evolutionary stage of a protostar when it comes to the formation of a disk.

VLA1623 is a triple non-coeval protostellar system \citep{murillo2013, chen2013} with a prominent outflow \citep{andre1990} located in $\rho$ Ophiuchus ($d \sim$ 120pc, \citealt{loinard2008}). VLA1623 is composed of three continuum sources: VLA1623A, a deeply embedded Class 0 source with no emission shortward of 24 $\mu$m; VLA1623B, a very cold and compact source separated 1.1$\arcsec$ from VLA1623A; and VLA1623W, a Class I source at a projected distance of 10" from VLA1623A. SMA observations of VLA1623 revealed that the easternmost component, VLA1623A, has a disk-like structure in C$^{18}$O and C$^{17}$O which was mimicked in continuum. Furthermore, VLA1623A's disk shows a velocity gradient characteristic of rotation. A simple eyeball analysis of the Position-Velocity diagrams of the C$^{17}$O and C$^{18}$O emission by \cite{murillo2013} suggested that the disk was rotationally supported, most likely exhibiting pure Keplerian rotation. They concluded that further analysis is required to distinguish the pseudo-disk $v \propto R^{-1}$ pattern from the rotationally supported $v \propto R^{-0.5}$ profile.

\begin{figure*}
\includegraphics[width=17cm]{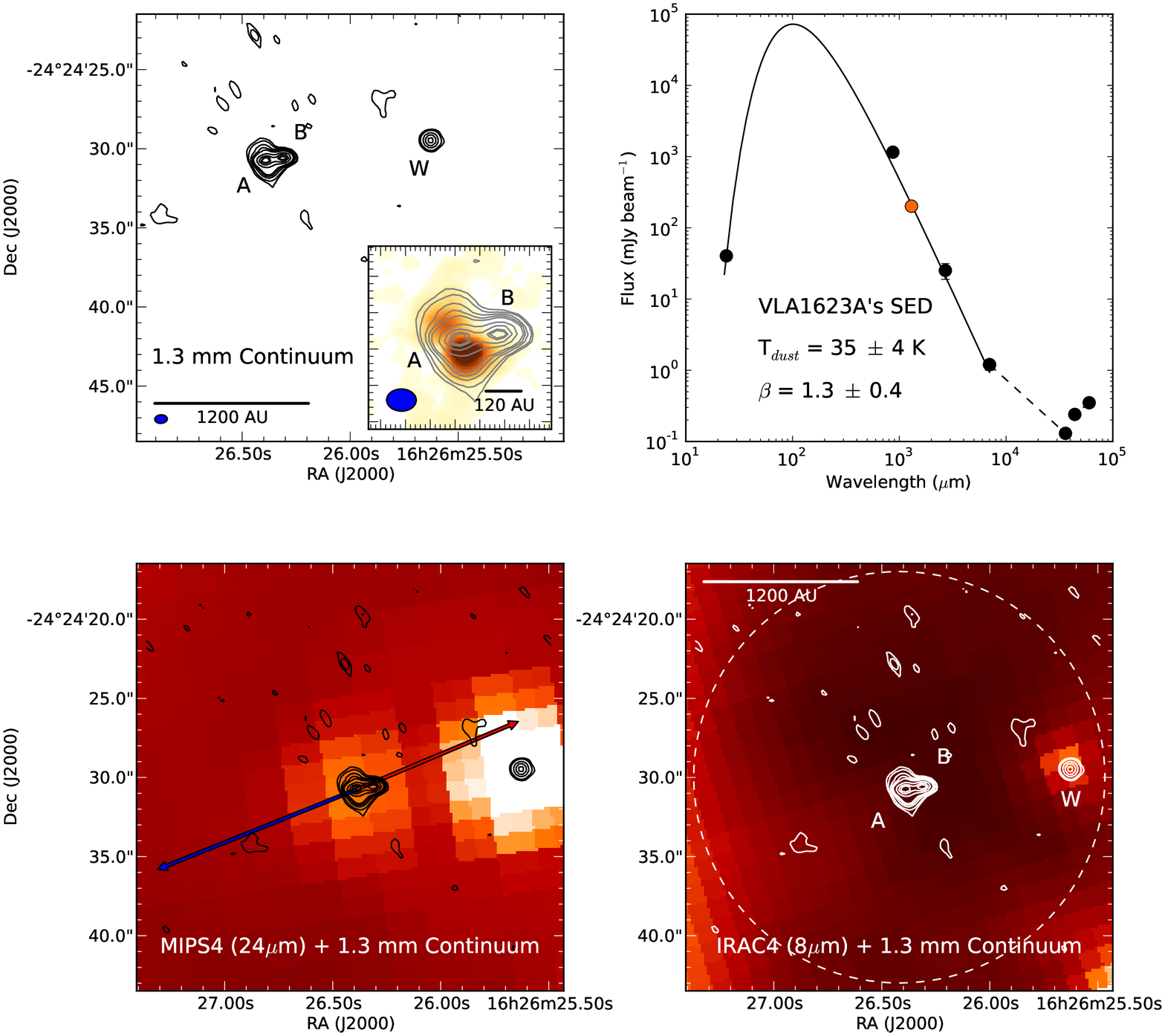}
\caption{VLA1623 in Continuum: \emph{Top left:} ALMA 1.3 mm continnum, three components are observed: VLA1623A, B and W. While B and W are observed to be compact, A shows an extended and flattened component. The inset shows a zoom-in of VLA1623A and B (black contours), overlaid with intensity integrated C$^{18}$O (2-1) (halftone).  Contours are in steps of 3$\sigma$, 5$\sigma$, 10$\sigma$, 15$\sigma$, 20$\sigma$, 40$\sigma$, 60$\sigma$ and 78$\sigma$, with $\sigma$=1 mJy beam$^{-1}$. \emph{Top right}: VLA1623A's SED; the orange point indicates VLA1623A's integrated continuum flux obtained from our ALMA observations. The solid line is the best fit of a single-temperature graybody fit, the resulting parameters are indicated in the figure. Flux uncertainties are usually smaller than the plot symbols. \emph{Bottom left}: ALMA 1.3 mm continuum contours overlaid over Spitzer MIPS1. Contours are the same as in the top left figure. Red and blue arrows indicate the red- and blueshifted large scale outflow direction. \emph{Bottom right}: ALMA 1.3 mm continuum contours overlaid over Spitzer IRAC4. Contours are the same as in the top left figure. The dashed circle indicates the field of view of the ALMA observations.}
\label{figcont}
\end{figure*}

In this paper, we present Atacama Large Millimeter/submillimeter Array (ALMA) Early Science Cycle 0 observations of VLA1623A in 1.3 mm continuum and C$^{18}$O emission. The results of the C$^{18}$O ALMA observations are consistent with previous Sub-Millimeter Array (SMA) C$^{18}$O observations in that they both show the disk-like structure and velocity gradient centered around VLA1623A. However, the sensitivity, spatial and velocity resolutions of the detection are significantly increased/improved due to ALMA's capabilities. In order to determine whether the C$^{18}$O emission is tracing a rotationally supported disk about VLA1623A, we perform Position-Velocity diagram analysis and models of the C$^{18}$O emission in order to determine the extent to which the observed structure is rotationally supported.

\section{Observations}
\label{secobs}
We observed VLA1623 (pointing coordinates $\alpha$=16:26:26.419 $\delta$=-24.24.29.988 J2000) with the Atacama Large Millimeter/submillimeter Array (ALMA) during the Early Science Cycle 0 period on April 8th, 2012. Observations were done in Band 6 (230 GHz) using the extended configuration, consisting of 16 antennae with a maximum baseline of $\sim$400m, for a total observing time of 1 hour and a duty cycle of 58\%. Calibration was done with 3C 279, 1733-130 and Titan for bandpass, gain and flux calibration, respectively. The spectral configuration was set-up to observe four molecular lines: DCO$^{+}$ (3-2), C$^{18}$O (2-1), N$_{2}$D$^{+}$ (3-2) and $^{12}$CO (2-1), in addition to continuum. No significant N$_{2}$D$^{+}$ emission was detected. In this paper we will only analyse and discuss the results of C$^{18}$O and continuum. The remaining lines will be discussed in future papers. The spectral configuration provided a velocity resolution of 0.0833 km s$^{-1}$ for C$^{18}$O.

Reduced data were received on June 15th, 2012. We re-did the data calibration and reduction using the standard pipeline for single pointing observations. Comparison of the delivered data and our recalibrated data shows consistency in continuum flux levels. The synthesized beam size is 0.79$\arcsec$ $\times$ 0.54$\arcsec$ for continuum and 0.79$\arcsec$ $\times$ 0.61$\arcsec$ for C$^{18}$O, providing enough resolution to resolve the continuum and line emission from each source.

We compare the ALMA 1.3 mm continuum flux with that obtained with SMA for the same wavelength reported in \citet{murillo2013}. To compare, VLA1623B's integrated flux is used since it was the most sensitive to the flux calibrations. We use the same region and task (CASA's imstat task) to measure the integrated flux for both datasets. For the SMA data, VLA1623B has an integrated flux of 96.8$\pm$0.7 mJy, while for the ALMA data, VLA1623B has an integrated flux of 96.3$\pm$0.7 mJy. This shows that both observations are consistent.

JCMT observations towards VLA1623 by \cite{jorgensen2002} detected C$^{18}$O (2-1) with a flux of 12.1 K km s$^{-1}$ (174 Jy km s$^{-1}$, with 14.46 Jy/K). Our C$^{18}$O (2-1) detection with ALMA has a flux of 13 Jy km s$^{-1}$. Thus, we recover $\sim$8\% of the total single dish detected flux.

\begin{figure*}
\includegraphics[angle=-90,width=\linewidth]{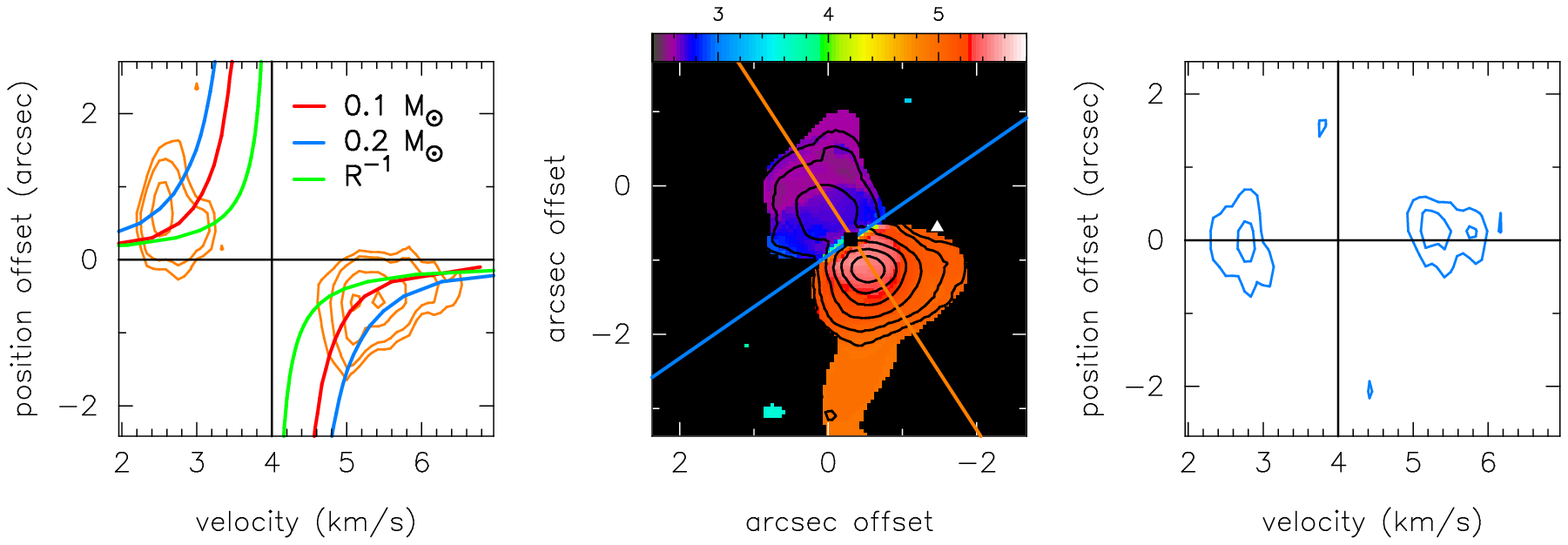}
\caption{Observed C$^{18}$O (2-1): \emph{Left}: Pure Keplerian rotation curves ($v \propto R^{-0.5}$, red and blue) and infall ($v \propto R^{-1}$, green) are overlaid on the PV diagram. This suggests that the emission may be rotationally supported with $M_{*}$ $\sim$ 0.1-0.2 M$_{\sun}$. \emph{Center}: C$^{18}$O velocity map (Moment 1, halftone) and intensity integrated (Moment 0, contours) maps. Contours are in steps of 3$\sigma$, 5$\sigma$, 10$\sigma$, 15$\sigma$, 20$\sigma$ and 25$\sigma$ with $\sigma$ = 13 mJy beam$^{-1}$. VLA1623A and B's positions are marked with a square and triangle, respectively. Orange and blue lines indicate the image-space PV diagram cuts at PA = 35$^{\circ}$ (left) and 125$^{\circ}$ (right), respectively. \emph{Right}: Lack of velocity gradient suggests no outflow contamination on the C$^{18}$O emission. In both PV diagrams contours are in steps of 3$\sigma$, 5$\sigma$, 10$\sigma$ and 15$\sigma$ where $\sigma$ = 19 mJy beam$^{-1}$ and the black lines indicate the systemic velocity and position of VLA1623A.}
\label{figmom}
\end{figure*}

\section{Results}
\label{secres}
\subsection{1.3 mm Continuum}
\label{seccont}
ALMA 1.3 mm continuum observations detect three continuum sources (Figure~\ref{figcont}, top left), in agreement with previous SMA observations \citep{murillo2013,chen2013}. The easternmost source, VLA1623A, shows an elongated and flattened morphology peaking in the center, with the elongation perpendicular to the outflow direction (Figure~\ref{figcont}, top left insert). VLA1623A is detected at 24 $\mu$m, but shows no emission shortward of 24 $\mu$m (Figure~\ref{figcont}, bottom row). VLA1623B, located to the west of VLA1623A and separated by 1.1$\arcsec$, is compact and shows no discernible infrared emission (Figure~\ref{figcont}, bottom row) suggesting that it is deeply embedded and cold. The westernmost source, VLA1623W ($\sim$10$\arcsec$ separation), is observed to be compact and dimmer than the other two sources in millimeter, but is bright in the infrared (Figure~\ref{figcont}, bottom row). Although VLA1623W is located at the edge of the field of view in our ALMA observations, it is a $\sim$20$\sigma$ detection and its flux and position coincide with previous detections of this source, thus we consider it a genuine detection. Peak and integrated fluxes of each source are listed in Table~\ref{tabcont}. Integrated fluxes for each source were obtained by integrating the continuum emission over a region the size of the source out to 3$\sigma$. In this paper, we will focus on VLA1623A, whose SED is shown in Figure~\ref{figcont}, top right.

\subsection{C$^{18}$O (2-1)} 
\label{secc18o}
The C$^{18}$O (2-1) emission towards VLA1623A shows an elongated and flattened structure perpendicular to the outflow direction, following the same shape as VLA1623A's 1.3 mm continuum emission (Figure~\ref{figcont}, top left insert, Figure~\ref{figmom}, center). Results and discussion of the C$^{18}$O emission towards VLA1623B and W are presented in Appendix~\ref{secW}.

The C$^{18}$O emission about VLA1623A exhibits a velocity gradient along the major axis, with blue-shifted material to the NE and red-shifted material to the SW in the velocity range of 2 to 6.5 km s$^{-1}$ (Figure~\ref{figmom}). The results of our ALMA observations presented here are consistent with SMA observations of C$^{18}$O (2-1) towards VLA1623A \citep{murillo2013}. In comparison, however, our ALMA observations have higher sensitivity (rms noise = 18 mJy beam$^{-1}$) and velocity resolution (0.0833 km s$^{-1}$) than previous SMA observations (rms noise = 42 mJy beam$^{-1}$ and velocity resolution = 0.275 km s$^{-1}$). ALMA's higher sensitivity and velocity resolution allow us to obtain a better view of the structure being traced and a deeper kinematical analysis. In addition, the high sensitivity allowed the detection of previously unknown filament-like features, located to the north and south of VLA1623A's C$^{18}$O disk structure (Figure~\ref{figmom}, center). These features may be the beginnings of a disk wind \citep{klaassen2013}. 

The observed velocity gradient along the major axis suggests rotation (Figure~\ref{figmom} left and center), as previously suspected from the SMA detection of C$^{18}$O. Furthermore, lack of a velocity gradient along the outflow axis indicates that the observed C$^{18}$O line emission is tracing only the envelope and/or disk structure (Figure~\ref{figmom} right). 

\begin{figure}
\includegraphics[width=\linewidth]{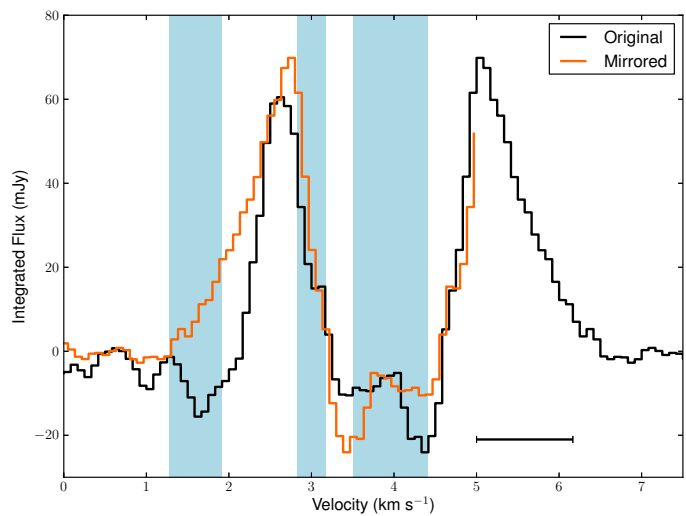}
\caption{C$^{18}$O (2-1) line profile. The unaltered spectral profile (black) is overlaid with a portion of the spectral profile mirrored about 4.0 km s$^{-1}$ (orange). Light blue rectangles mark the widths obtained from the best fit thin disk model (Table~\ref{tabfg}) of the foreground material (1.6 and 3 km s$^{-1}$) and the envelope (4.0 km s$^{-1}$). The horizontal line between 5 and 6.5 km s$^{-1}$ shows the velocity range used for the UV-space PV diagram.}
\label{figspec}
\end{figure}

\begin{figure}
\includegraphics[width=\linewidth]{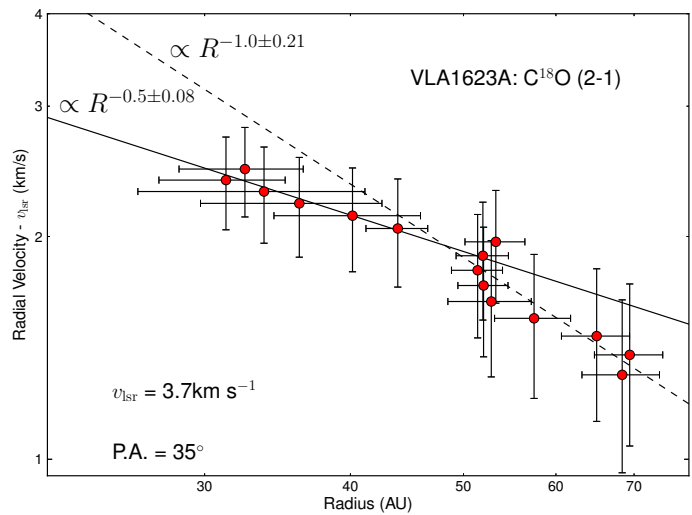}
\caption{UV-space PV diagram of C$^{18}$O. $v_{\rm lsr}$ = 3.7 km s$^{-1}$ was used. The red points indicate the red-shifted ($v$ > 5.0 km s$^{-1}$) emission. The velocity range covered in this diagram is indicated in Figure~\ref{figspec}. Blue-shifted emission ($v$ < 2.8 km s$^{-1}$) is not included in the diagram since it is greatly affected by foreground emission (see Figure~\ref{figspec} and Section~\ref{secc18o}). The points are fit with a power law of the form $v = aR^{n}$. The solid line shows the fit to the high velocity points ($v$ > 5.7 km s$^{-1}$) while the dashed line shows the fit to the low velocity points ($v$ < 5.7 km s$^{-1}$). This suggests that the pure Keplerian disk extends to a radius of 50 AU.}
\label{figuvpv}
\end{figure}

The $\rho$ Ophiuchus region is known to have several layers of foreground clouds along the line of sight \citep{loren1989}, as evidenced by studies towards core B and E in $\rho$ Ophiuchus \citep{loren1989, boogert2002, kempen2009}. However, there are no current studies on the foreground of VLA1623 ($\rho$ Ophiuchus core A). From JCMT observations towards VLA1623, it is difficult to determine the position of absorbing material given the dense and active region \citep{jorgensen2004}. Determining the foreground over a large area from single dish data requires a more in depth study and is outside the scope of this paper. Thus to estimate the positions of the absorbing material, we compare the original and mirrored spectra (Figure~\ref{figspec}). From line observations, \cite{mardones1997} and \cite{yu1997} report VLA1623's systemic velocity $v_{\rm lsr}$ to be between 3.4 to 3.8 km s$^{-1}$. Mirroring the spectra about this range of v$_{\rm lsr}$ does not produce a symmetric spectra, however. We instead find that the C$^{18}$O spectra is better mirrored about 4.0 km s$^{-1}$, which would suggest that VLA1623's $v_{\rm lsr}$ might be 4.0 km s$^{-1}$. We then consider the full range of $v_{\rm lsr}$ = 3.4 to 4.0 km s$^{-1}$ in this work. Regarding the absorbing material, it is clear from Figure~\ref{figspec} that there is absorption at velocities near 2 and 3 km s$^{-1}$, as well as absorption towards the systemic velocity. The absorption at the systemic velocity is either due to absorption caused by VLA1623's outer envelope or resolved out emission. On the other hand, the absorptions near 2 and 3 km s$^{-1}$ are more likely to be due to foreground clouds. These foreground clouds "eat out" some of the blueshifted emission, causing the observed red- and blueshifted C$^{18}$O emission to seem asymmetric towards VLA1623A.

\begin{figure*}
\includegraphics[angle=-90]{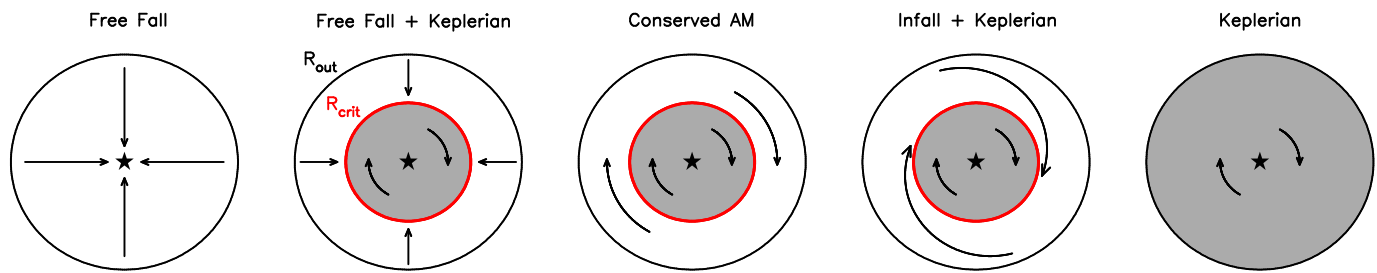}
\caption{Cartoon illustrating the velocity structure distribution in the disk, shown here face on, for each model examined.}
\label{figcar}
\end{figure*}

\section{Analysis} 
\label{secanal}
\subsection{Position-Velocity Diagrams}
\label{secpv}
The first and most common way of analysing the kinematics of line emission is through Position-Velocity (PV) diagrams. Here we perform image-space and UV-space PV diagrams. Image-space PV diagrams project the line emission 3D data cube into a 2D plane along a cut, defined by position angle (PA) and source position, over the structure of interest. UV-space PV diagrams, on the other hand, are constructed by fitting, channel by channel, the UV visibilities of the line emission to find the peak position in each channel \citep{lommen2008,jorgensen2009}. The peak positions of each channel are then rotated to the position angle along the velocity gradient, projected onto Position-Velocity space and fit with a power law in log-log space. The resulting PV diagram and power law fit provide insight into the kinematic structure of the line emission. 

\begin{figure*}
\includegraphics[angle=-90,scale=1.352]{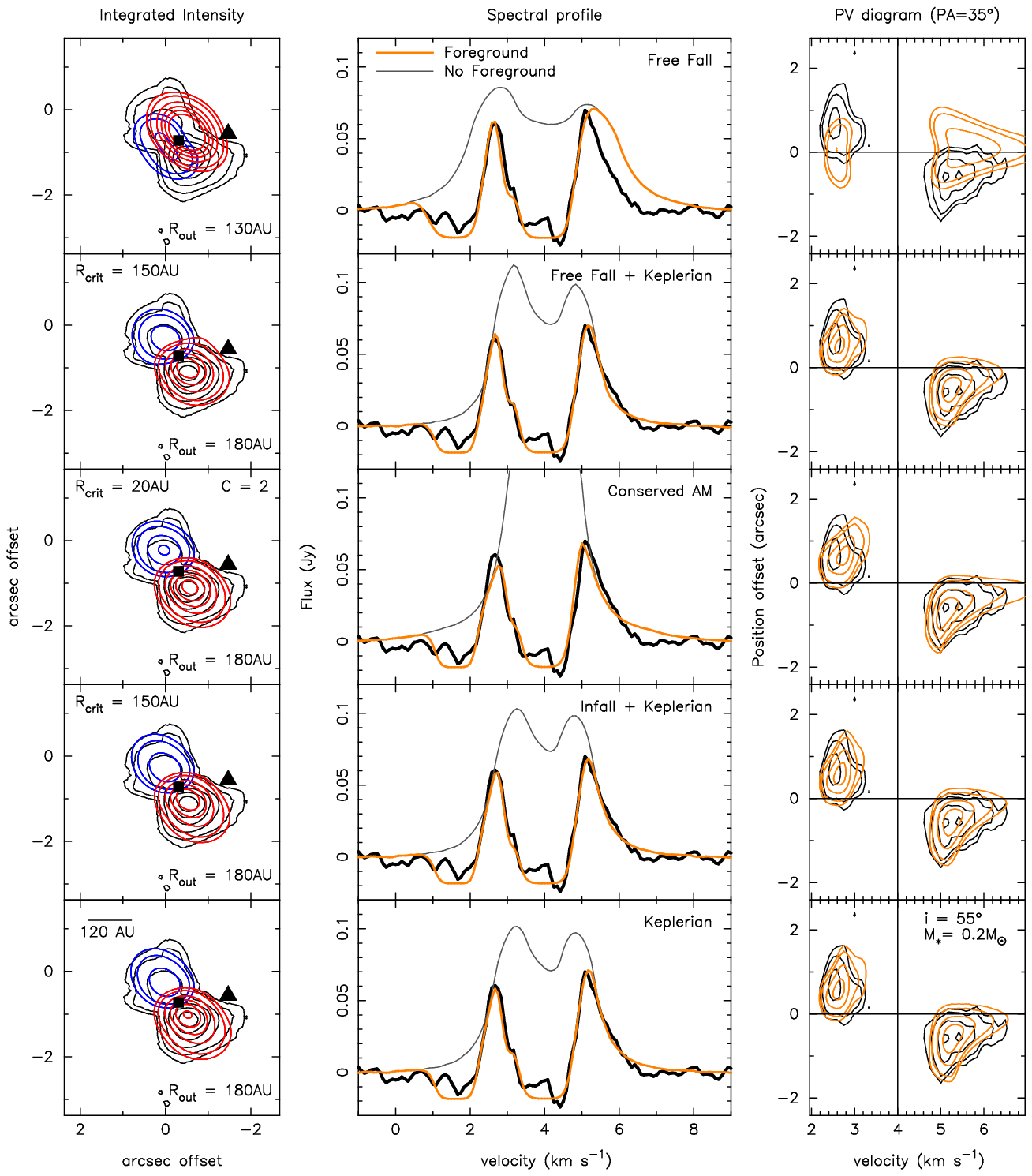}
\caption{Comparison of C$^{18}$O observations (black lines) with a thin disk model (colored lines) that includes foreground components. The left column shows moment 0 maps with the red and blue lines showing red and blueshifted emission, respectively. Source positions are marked as in Figure~\ref{figmom}. The middle column compares the spectral profiles with and without foreground. The right column presents the image-space PV diagrams. The models shown here have two foreground components (at 1.6 and 3.0 km s$^{-1}$) and the resolved out envelope component (at 4.0 km s$^{-1}$). Contours both for observations and models are the same as those in Figure~\ref{figmom}. $M_{*}$ and $i$ for all models are shown in the last panel.}
\label{figmod}
\end{figure*}

The C$^{18}$O velocity integrated map (Figure~\ref{figmom}, center) shows a velocity gradient along the major axis (PA = 35$\pm$15$^{\circ}$) of the structure, perpendicular to the outflow. The image-space PV diagram along the structure's major axis (Figure~\ref{figmom}, left) is characteristic of a rotating structure around a 0.1 to 0.2 M$_{\sun}$ object, with pure infall clearly providing a worse fit. The rotation curves in Figure~\ref{figmom} (left panel) do not account for inclination effects.

Figure~\ref{figuvpv} shows the UV-space PV diagram of the C$^{18}$O emission. The blue-shifted emission is excluded from the diagram since it is greatly affected by foreground emission (Figure~\ref{figspec}, also see Section~\ref{secc18o}) at higher velocities. Thus we only fit the red-shifted points, with velocities greater than 5.0 km s$^{-1}$. The velocity range of points used is shown in Figure~\ref{figspec}. The data points are fit with a power law of the form $v = aR^{n}$ and we find that the UV-space PV diagram is independent of the inclination angle of the structure. The entire range of possible $v_{\rm lsr}$ (3.4 to 4.0 km s$^{-1}$) was fit. The results of the fit for any $v_{\rm lsr}$ are within the error of the fit presented here, which uses the mean $v_{\rm lsr}$, 3.7 km s$^{-1}$. Thus we find that the high velocity points ($v$ > 5.7 km s$^{-1}$) are well fit by $v \propto R^{-0.5 \pm 0.08}$, giving a central protostellar mass of $M_{*}$ of 0.22$\pm$0.02 M$_{\sun}$. On the other hand the low velocity points (5 km s$^{-1}$ < $v$ < 5.7 km s$^{-1}$) are well fit by $v \propto R^{-1}$. This suggests that there is a turn over point at 50 AU, where the inner parts, i.e.
$R \leq$ 50 AU, are under pure Keplerian rotation and the outer parts are under infall. In the following section, we will argue that the lower velocity points are affected by optical depth and absorption, and that $R$ can be larger.

\subsection{Modelling of C$^{18}$O (2-1)}
\label{secmod}
Based on the results from the PV diagrams, which suggest the presence of a rotationally supported component in the observed C$^{18}$O emission, we proceed to further investigate the structure and its kinematics using an analytical thin disk model with the addition of absorbing foreground material. The model does not include radiative transfer or excitation since the goal is to study the kinematics and structure of the observed disk. The input parameters include the outer radius $R_{\rm out}$ of the disk, source position, position angle P.A. and inclination $i$ of the disk, the mass of the central source $M_{*}$ and the temperature and column density gradients of the disk surface. Generated maps are convolved to the observed clean beam. The model output are in the form of position-position-velocity (xyv) data cubes in FITS format.

For all of the models we fix the central protostellar mass $M_{*}$ = 0.2 M$_{\sun}$, which was obtained from the PV diagram analysis (see Section~\ref{secpv}), and the position angle to 35$^{\circ}$, as these parameters are well constrained from the PV diagrams (see Section~\ref{secpv} and Figures~\ref{figmom} and~\ref{figuvpv}). We set $v_{\rm lsr}$ = 4.0 km s$^{-1}$, which is the symmetry axis of the spectrum. The distance is set to 120 pc and the source position is set slightly offset by 0.02$\arcsec$ $\times$ 0.04$\arcsec$ to the SW from VLA1623A's position to match the center of the C$^{18}$O emission, i.e. the rotation axis. 

Other free parameters in the model are line width $v_{\rm width}$, inclination $i$ and outer radius $R_{\rm out}$. These values are constrained to best fit the observations. The attempted value ranges for each parameter are: 0.1 km s$^{-1}$ $\leq$ $v_{\rm width}$ $\leq$ 1.0 km s$^{-1}$; 0$^\circ$ (face-on) $\leq$ $i$ $\leq$ 90$^\circ$ (edge-on); 100 AU $\leq$ $R_{\rm out}$ $\leq$ 200 AU. The best fit for almost all models was obtained with $v_{\rm width}$ = 0.2 km s$^{-1}$, $i$ = 55$^\circ$ and $R_{\rm out}$ = 180 AU, except for the Free falling disk model, where $R_{\rm out}$ = 130 AU. Since we do not have a way to constrain the temperature and column density gradients, we assume the observed emission is optically thin, and set the density as a constant and adjust the temperature gradient to match the observed spectral profile.

As discussed in Section~\ref{secc18o}, we suspect the presence of absorbing material along the line of sight of VLA1623 which would affect the observed emission and spectral profile, and thus influence the model fitting. From the data presented here we can not be completely certain on the characteristics of the absorbing material. Thus, we adjust the model velocities, peak opacity and widths of the foreground clouds aiming to best fit the observed spectral profile. We assume that the foreground clouds only absorb with opacity following a Gaussian-like profile of the form
\begin{equation}
\tau(v) \propto \exp\left(-0.5\left(\frac{v - v_{0}}{\sigma}\right)^2\right).
\end{equation} 

By adjusting the absorbing material parameters to fit the observed spectral profile, we find that the best results are obtained by introducing three absorbing layers, one corresponding to the envelope at 4.0 km s$^{-1}$, and two corresponding to possible foreground clouds at 1.6 and 3.0 km s$^{-1}$. The characteristics of the absorbing material for each model are listed in Table~\ref{tabfg}.

\begin{figure}
\includegraphics[angle=-90,width=\linewidth]{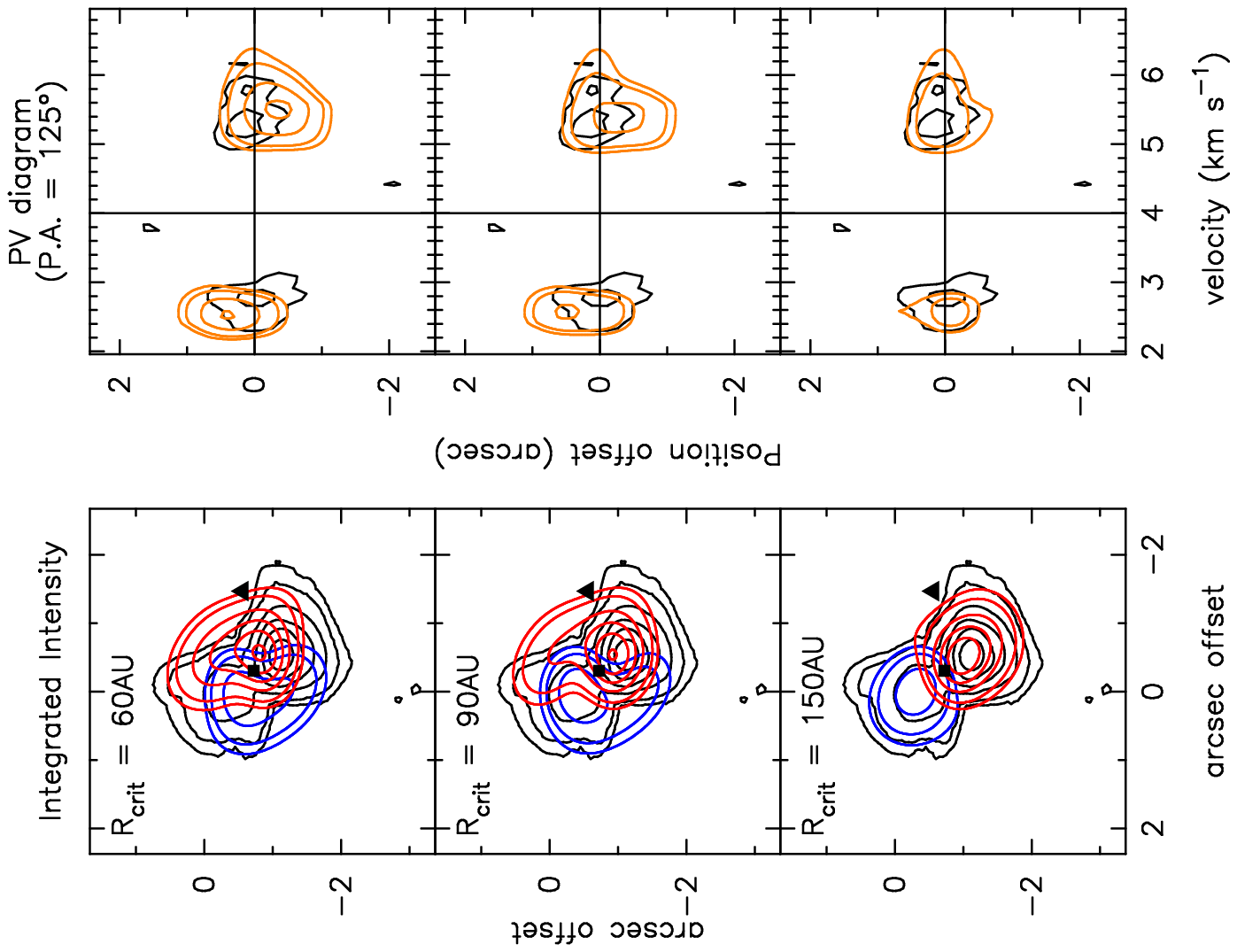}
\caption{Free fall plus Keplerian case for three different $R_{\rm crit}$. The best fit model (color) to the data (black) is obtained as $R_{\rm crit}$ approaches $R_{\rm out}$. Right panels show PV diagram along outflow direction (P.A. = 125$^{\circ}$). All contours and the model at $R_{\rm crit}$ = 150 AU are the same as in Figure~\ref{figmod}.}
\label{figmodmix}
\end{figure}

\begin{figure}
\includegraphics[angle=-90,width=\linewidth]{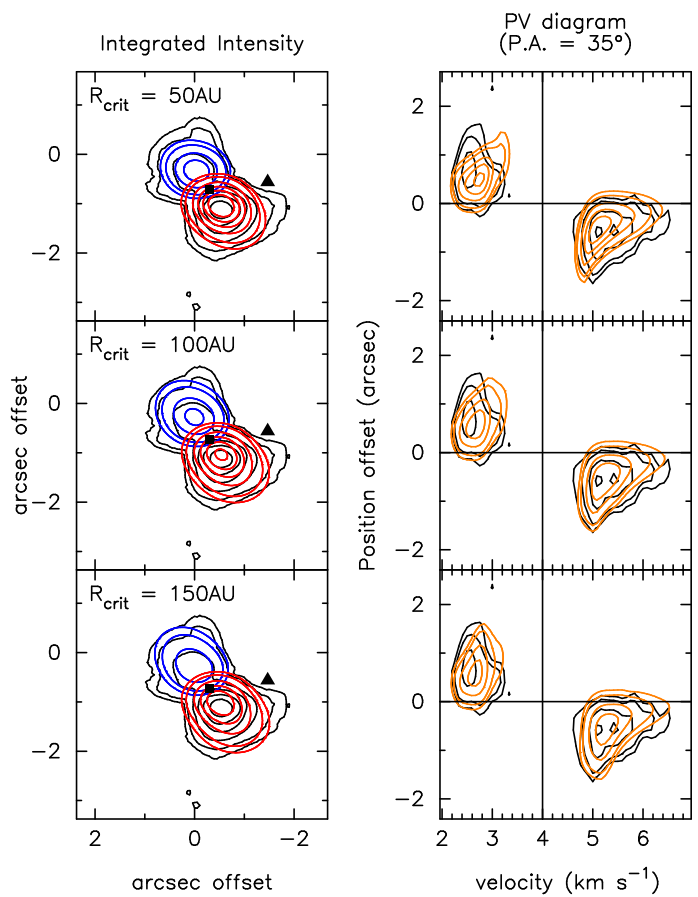}
\caption{Infall plus Keplerian case for three different $R_{\rm crit}$. The best fit model (color) to the data (black) is obtained as $R_{\rm crit}$ approaches $R_{\rm out}$. Right panels show PV diagrams along the disk's major axis (P.A. = 35$^{\circ}$). All contours and the model at $R_{\rm crit}$ = 150 AU are the same as in Figure~\ref{figmod}.}
\label{figmodpv}
\end{figure}

\begin{table*}
\caption{Best Fit Thin Disk Model Results}
\label{tabmod}
\centering
\begin{tabular}{c c c c c c}
\hline \hline
Parameter & Free Fall & Free Fall + Keplerian & Conserved AM & Infall + Keplerian & Keplerian \\
\hline
$i$ (degrees) & 55 & 55 & 55 & 55 & 55 \\
$R_{\rm out}$ (AU) & 130 & 180 & 180 & 180 & 180 \\
$R_{\rm crit}$ (AU) & ... & 150 & 20 & 150 & ... \\
\hline
\end{tabular}
\end{table*}

\begin{table*}
\caption{Absorbing material for best fit thin disk model}
\label{tabfg}
\centering
\begin{tabular}{l c c c c c}
\hline \hline
 & Free Fall & Free fall + Keplerian & Conserved AM & Infall + Keplerian & Keplerian \\
\hline
\emph{Envelope} & & & & & \\
  ~~velocity (km s$^{-1}$) & 4.0 & 4.0 & 4.0 & 4.0 & 4.0 \\
  ~~opacity & 7.0 & 7.0 & 7.0 & 7.5 & 7.0 \\
  ~~width (km s$^{-1}$) & 0.9 & 0.99 & 0.99 & 0.99 & 0.99 \\
\emph{Foreground 1} & & & & & \\
  ~~velocity (km s$^{-1}$) & 1.6 & 1.6 & 1.6 & 1.6 & 1.6 \\
  ~~opacity & 5.0 & 5.0 & 5.0 & 5.0 & 5.0 \\
  ~~width (km s$^{-1}$) & 0.89 & 0.64 & 0.62 & 0.64 & 0.64 \\
\emph{Foreground 2} & & & & & \\
  ~~velocity (km s$^{-1}$) & 3.0 & 3.0 & 3.1 & 3.05 & 3.0 \\
  ~~opacity & 0.39 & 0.29 & 0.29 & 0.29 & 0.29 \\
  ~~width (km s$^{-1}$) & 0.39 & 0.34 & 0.34 & 0.34 & 0.34 \\
\hline
\end{tabular}
\end{table*}

In order to study the kinematical structure of the disk, we model five cases: (1) free falling disk, (2) free falling outer disk plus inner Keplerian disk, (3) Conservation of Angular Momentum (AM), (4) infalling ($v \propto R^{-1}$) outer envelope plus inner Keplerian disk, and (5) pure Keplerian disk. For cases 2, 3 and 4, we define a critical radius $R_{\rm crit}$ parameter (1 AU $\leq$ $R_{\rm crit}$ $<$ $R_{\rm out}$) which defines the radius at which the transition from one velocity structure to the other occurs. Figure~\ref{figcar} illustrates the velocity structure distribution of each of the modelled cases. In Section~\ref{secmodres} we briefly describe each model and the results obtained.

Figure~\ref{figmod} presents the best fitting results for the five cases modelled. We overlay the observed and modelled spatial distribution, spectral profile and image-space PV diagram. Model spectral profiles, are shown both with and without foreground absorption. The image-space PV diagram for each model was constructed along the disk's major axis. The best fit parameters obtained for each of the models are shown in Figure~\ref{figmod} and listed in Table~\ref{tabmod}. To determine the best fit model of all five cases examined, we first compared each model and observations by eye and then through the residuals obtained by subtracting the channel maps of the observations and model. We deemed the model with the least residual the best fit (discarding the filaments to the north and south). Appendix~\ref{seccomp} shows the channel map for each model, compared with observations and the corresponding residual.

\subsection{Model Results}
\label{secmodres}
\textbf{Free falling disk:} At the very early stages of star formation, the envelope of a protostar is thought to be under free fall collapse. Thus, if there was any disk-like structure around the protostar at this stage, its motion would be that of free fall as well, i.e. a free falling disk (Figure~\ref{figcar}). In our model, we study a free falling disk structure, where all motions are confined to the plane of the disk. The velocity structure within the disk structure is described as:
\begin{equation}
v^{r}_{\rm ff} = \sqrt{\frac{2GM_{*}}{R}}
\label{eqff}
\end{equation}

In Figure~\ref{figmod}, the top row shows the thin disk model under free fall. In Figure~\ref{figresff} the channel map of the model and the residual are shown. It is clear that the pure free fall case does not fit the observed C$^{18}$O emission, since the velocity gradient is along the outflow direction, which is not the case in the observations. In addition, the wings in the spectral profile of this model are broad compared to the observed spectral profile.

\textbf{Free fall plus Keplerian rotation:} this case represents the formation of a rotationally supported disk in a young protostar whose outer disk is under free fall collapse (Figure~\ref{figcar}). To describe the velocity structure transition, we use the parameter $R_{\rm crit}$, where $R < R_{\rm crit}$ produces

\begin{equation}
v^{\phi}_{\rm rot} = \sqrt{\frac{GM_{*}}{R}}
\label{eqkep}
\end{equation}
and $R \geq R_{\rm crit}$ is described by Equation~\ref{eqff}.

For this case, we experiment with different critical radii $R_{\rm crit}$ to determine if the observed emission is a combination of Keplerian rotation and free fall (Figure~\ref{figmodmix}). We find that the free fall region ($R > R_{\rm crit}$) is clearly visible unless $R_{\rm crit}$ approximates $R_{\rm out}$. Similar to the pure free fall case, the outer free fall region produces a velocity gradient along the outflow direction, which is visible in the PV diagram (Figure~\ref{figmodmix}), however we do not observe any velocity gradient along the outflow direction in the C$^{18}$O structure (Figure~\ref{figmom}, right). In addition, the C$^{17}$O (3-2) emission observed with SMA \citep{murillo2013} shows a velocity gradient indicative of rotation, thus the outer part of the observed structure is not under free fall. Figures~\ref{figmodmix} and \ref{figresffk} compare this model and the observations in detail, which show that only at $R_{\rm crit}$ = 150 AU does the model closely approximate the observations.

\textbf{Conserved Angular Momentum (AM):} this case examines whether VLA1623A's C$^{18}$O structure is well described by rotation with conserved angular momentum (Figure~\ref{figcar}), for example if the initial angular momentum of the cloud would be conserved during collapse. The velocity structure in the inner region  $R < R_{\rm crit}$ is given by Equation~\ref{eqkep} while at $R \geq R_{\rm crit}$ it is described by:
\begin{equation}
v^{\phi}_{\rm AM} = C \sqrt{\frac{GM_{*}}{R_{\rm crit}}} \frac{R_{\rm crit}}{R}
\label{eqsb}
\end{equation}
where C is a constant indicating the increase in the Keplerian velocity at $R_{\rm crit}$.

For this model, we find $C$ = 2 with $R_{\rm crit}$ = 20 AU provides the closest fit to the observations. In comparison to the other cases examined, this model produces broader wings and higher velocities at smaller radii, effects visible in the spectral profile and the PV diagram, respectively (Figure~\ref{figmod}). Additionally, from the spectral profile shown in Figure~\ref{figmod}, it is seen that this model does not correctly reproduce the blue-shifted peak. Changing the $C$ parameter does not resolve this issue, and changing the line width additionally broadens the wings. Furthermore, even though from the intensity integrated map the spatial distribution appears to fit the observations, inspection of the channel map residuals show that the model does not quite agree with the observations (Figure~\ref{figrescam}). Thus we determine that this scenario does not match the observations.

\begin{table*}
\caption{Class 0 sources with disk}
\label{tabcomp}
\centering
\begin{tabular}{c c c c c}
\hline \hline
Parameter & NGC1333 IRAS4A2\tablefootmark{a} & L1527 & VLA1623A & Ref.\\
\hline
$i$ (degrees) & 10.7 & $\sim$85 & 55 & 1,2,3 \\
$R_{\rm out}$ (AU) & 310 & 90 & 180 & 1,2,3 \\
$M_{*}$ (M$_{\sun}$) & 0.08$\pm$0.02 & 0.19$\pm$0.04 & 0.22$\pm$0.02 & 1,2,3 \\
$M_{\rm env}$\tablefootmark{b} (M$_{\sun}$) & 5.6 & 0.9 & 0.8 & 4,5 \\
$M_{*}$/$M_{env}$ & 0.01 & 0.2 & 0.28 & ... \\
$T_{\rm bol}$ (K) & 51 & 44 & 10 & 1,5,3 \\
$L_{\rm bol}$ ($L_{\sun}$) & 1.9$\pm$0.9 & 1.97 & 1.1$\pm$0.2 & 1,2,6 \\
$L_{\rm submm}$/$L_{\rm bol}$\tablefootmark{c} (\%) & 3.6 & 0.8 & 1.2 & ... \\
\hline
\end{tabular}
\\
\tablebib{
(1)~\citet{choi2010}; (2)~\citet{tobin2012}; (3)~this work; (4)~\citet{froebrich2005}; (5)~\citet{kristensen2012}; (6)~\citet{murillo2013}
}
\raggedright
\tablefoottext{a}{$T_{\rm bol}$ and $L_{\rm bol}$ parameters are for NGC1333 IRAS4A, \cite{choi2010} assumes IRAS4A2 contributes half of the $L_{\rm bol}$.}
\tablefoottext{b}{$M_{\rm env}$ is the total envelope mass of each system.}
\tablefoottext{c}{$L_{\rm submm}$/$L_{\rm bol}$ > 0.5\% is characteristic of Class 0 source \citep{froebrich2005, andre1993}.}
\end{table*}

\textbf{Infall plus Keplerian rotation:} similar to the free fall plus Keplerian case, but with the outer part of the envelope infalling onto the rotationally supported disk (Figure~\ref{figcar}). We consider this model a possibility from the results of the UV-space PV diagram (Figure~\ref{figuvpv}) which shows an inner region under Keplerian rotation and an outer region of infall, with a critical radius of $\sim$50 AU. Thus, for the inner rotating region (i.e. $R < R_{\rm crit}$) the velocity structure is described by Equation~\ref{eqkep}, and for the outer infalling region ($R \geq R_{\rm crit}$) it is given by:
\begin{equation}
v^{r}_{\rm infall} = \sqrt{\frac{GM_{*}}{R_{\rm crit}}} \frac{R_{\rm crit}}{R}
\label{eqinfall}
\end{equation}

As in the free fall plus Keplerian case, we run the model with different critical radii $R_{\rm crit}$ to determine how far does each of the velocity structures extend (Figure~\ref{figmodpv}). Similar to the free fall plus Keplerian case, we find that the best fit occurs as $R_{\rm crit}$ approaches $R_{\rm out}$, with the best fit having $R_{\rm crit}$ = 150 AU (Figures~\ref{figmod} and ~\ref{figmodpv}). The discrepancy between the $R_{\rm crit}$ obtained from the modelling (150 AU) and that of the UV-space PV diagram from observations (50 AU) may be due to the low velocity points being greatly affected by optical depth and absorption of the envelope. Keplerian rotation out to at least 150 AU represents a best fit scenario for the observed C$^{18}$O emission. Because of the large $R_{\rm crit}$, it is difficult to distinguish this scenario from the following case. In Figure~\ref{figresink} the channel map comparison and residual between this model and the observations are shown.

\textbf{Pure Keplerian rotation:} in this case we model a Keplerian disk around a central protostar (Figure~\ref{figcar}). This velocity structure has been observed starting in Class I protostars and is common among Class II protostars. The velocity structure is given by Equation~\ref{eqkep}. 

We find this case to fit well the observed C$^{18}$O emission, as the model agrees with the observations spatially, in the spectral profile and in the image-space PV diagram (Figure~\ref{figmod}). However it is difficult to determine whether this scenario is considerably better than the infall plus Keplerian case. We further discuss this in the next section. In Figure~\ref{figreskep} the channel map comparison and residual between this model and the observations are shown.

\section{Discussion}
\label{secdis}
Our ALMA observations of VLA1623 reveal that the Class 0 component of this system, VLA1623A, has a disk structure in C$^{18}$O (2-1) with an outer radius of 180 AU. Thin disk modelling with the addition of foreground indicates that the disk structure is most certainly Keplerian out to a radius of 150 AU. For the outer 30 AU, it is uncertain whether the velocity structure may be under infall ($v \propto R^{-1}$) or pure Keplerian rotation. This uncertainty is due to the low S/N ratio (5$\sigma$) in the outer edges of the observed disk structure, while the inner regions of the disk have a higher S/N ratio (10 to 25$\sigma$), thus a better and more reliable fit is obtained in the inner parts of the observed structure. The presence of such a large Keplerian disk raises the question of the formation of disks in the early stages of protostellar evolution. 

Regardless of the model, neither the infall plus Keplerian case nor the pure Keplerian case reproduce the weak filament-like structures to the north of the blueshifted emission (v = 2.3 to 2.47 km s$^{-1}$) and south of the redshifted emission (v = 4.8 to 4.97 km s$^{-1}$) (Figure~\ref{figmom}, ~\ref{figresink} and ~\ref{figreskep}). These structures may be produced by a disk wind \citep{klaassen2013} or material entrained by the outflow.  

The idealized non-magnetized isothermal sphere collapse model of \cite{terebey1984} suggests that disks may form during the early protostellar stages. As previously mentioned, under these conditions the disk's centrifugal radius $R_{\rm c}$ is expected to grow proportional to $t^{3}$, where $t$ denotes the time since collapse. To calculate the time since collapse for $R_{\rm c}$ = 150 AU, we use the equation from \cite{belloche2013}
\begin{equation}
R_{\rm c}({\rm AU}) = 39 \left(\frac{\Omega}{4 \times 10^{-14}~{\rm rad~s^{-1}}}\right)^{2} \left(\frac{a}{0.2 ~{\rm km~s^{-1}}}\right) \left(\frac{m_{*+d}}{1~{\rm M_{\sun}}}\right)^{3}
\end{equation}
where $\Omega$ is the initial cloud core rotation rate, $a$ is the sound speed and $m_{*+d}$ = 0.975$\frac{a^3}{G}t$. Assuming VLA1623 has a rotation rate $\Omega$ = 4 $\times$ 10$^{-14}$ rad s$^{-1}$ and the sound speed in the core is of $a$ = 0.2 km s$^{-1}$, we obtain $t$ = 8.5 $\times$ 10$^{5}$ yr for a centrifugal radius of 150 AU. However, the old age obtained through this method is inconsistent with the expected age from $T_{\rm bol}$ \citep{ladd1998} and the estimated Class 0 lifetime obtained by \cite{evans2009}, which is of the order of 10$^{5}$ yr, although they conceded that for $\rho$ Ophiuchus the lifetime is of 4 $\times$ 10$^{4}$ yr, an order of magnitude lower than the above calculated age. Furthermore, based on outflow observations, VLA1623's dynamical timescale is between 0.2 and 2.5 $\times$ 10$^{4}$ yr \citep{andre1990, nakamura2011}. This indicates that there are other factors which enhance the formation of the disk. Possible factors that can influence disk formation may be fragmentation, turbulence or the misalignment of the magnetic field and rotation axis. On the first factor, fragmentation, there is little work on how fragmentation can enhance or hinder disk formation, with most work focusing on how a disk fragments. However, there may be a possible relation given that two of the three Class 0 sources reported of having rotationally supported disks are confirmed multiples (see Section~\ref{secintro} and below). The introduction of turbulence and its effect on disk formation from low to high masses has been studied by \cite{seifried2013} finding that turbulence can encourage disk formation even when strong magnetic fields are present. \cite{nakamura2011} studied the outflow generated turbulence in $\rho$ Ophiuchus' main cloud, concluding that outflows can significantly influence the dense cores. However, from \cite{nakamura2011} or this work, there is not enough information to determine the degree of influence that turbulence has on the formation of VLA1623A's disk. Hence at present we can neither further examine nor rule out the role of turbulence. The last factor, magnetic field misalignment, is considered in-depth below.

Magnetic fields are expected to influence the formation of protostellar disks in the early stages of protostar formation. It it thus of interest to look into the magnetic field configuration of VLA1623. \cite{holland1996} observed 800 $\mu$m polarization with the JCMT, finding a 2\% polarization and a magnetic field perpendicular to the outflow direction. \cite{hull2013} also observed the field to be perpendicular to the outflow down to 2" resolution with CARMA observations. \cite{murillo2013} carried out polarization observations with the SMA (compact configuration, resolution $\sim$ 1.5$\arcsec$) but found no significant detection. From these results we infer that the magnetic field towards VLA1623 is not aligned with the rotation axis of VLA1623A's disk, and the field strength may be low. \cite{krumholz2013} find that rotationally supported disks should form as early as the Class 0 stages with sizes of 100 AU or larger when the magnetic field direction and disk's rotational axis misalignment are large and the magnetic field strength is low. The presence of a fairly large rotationally supported disk around VLA1623A is consistent with \cite{krumholz2013}'s predictions, given VLA1623A's misalignment of 84$^{\circ}$ between magnetic field direction and rotational axis \citep{hull2013}. In addition, the disk size is sensitive to the initial cloud density profile, with centrally concentrated profiles favoring larger disks. Moreover the discrepancy between magnetized models of disk formation and VLA1623A's disk may be due to the sink particle parameters used to represent the forming protostar in the models \citep{machida2013}.

Two other Class 0 sources have been reported as presenting a Keplerian disk structure: NGC1333 IRAS4A2 and L1527. The characteristics of these sources are listed in Table~\ref{tabcomp} along with VLA1623A's characteristics. It should be noted that the parameters for NGC1333 IRAS4A2 are obtained under the assumption that this source contributes half of the bolometric luminosity and that both sources in the NGC1333 IRAS4A binary have the same bolometric temperature. Thus, in comparisson, VLA1623A is noticeably younger than L1527, based on the bolometric temperature and luminosity ratio. On the other hand, we cannot certainly determine the relative evolutionary stages of VLA1623A and NGC1333 IRAS4A2. We can assume, however, based on the bolometric temperature that VLA1623A is younger than NGC1333 IRAS4A2. It must be noted, though, that the inclination angle of a protostar can affect the calculated parameters and thus affect the derived evolutionary parameters. Given that all three sources have different inclination angles, this may well affect the comparison. However, it is very possible that VLA1623A's disk is the youngest disk among the three sources, given that VLA1623A is still deeply embedded and shows no emission shortward of 24$\mu$m.

\section{Conclusions}
\label{seconc}
We have presented ALMA Cycle 0 Early Science Band 6 extended configuration observations of C$^{18}$O (2-1) and continuum towards VLA1623. From these observations, we find three continuum sources which are consistent with previous observations and C$^{18}$O emission centered at VLA1623A with signatures of rotation and an outer radius of 180 AU. Through PV diagram analysis and modelling of the observed C$^{18}$O, we determine that the emission is tracing a Keplerian disk out to 150 AU around a 0.2$\pm$0.02 M$_{\sun}$ protostar. The weak magnetic field and its misalignment with the disk's rotational axis may have increased the chances of disk formation at such an early stage. There may also be the possibility that fragmentation played a role in the early disk formation, but it is unclear from the results obtained here. Comparison of evolutionary indicators of VLA1623A with those of the other Class 0 sources reported of showing indications of Keplerian disks suggest that VLA1623A's disk may be the youngest among the Class 0 Keplerian disks. Our results show that disks, and more precisely rotationally supported or Keplerian disks, can be formed in the Class 0 stage of protostellar evolution even with fairly large radii. However our results also hint at the fact that the environmental factors play a larger role than evolutionary stage in the formation of disks.

\begin{acknowledgements}
This paper makes use of the following ALMA data: ADS/JAO.ALMA\# 2011.0.00902.S which was obtained by N.M.M. while was a Master student at National Tsing Hua University, Taiwan, under the supervision of S.P.L.. ALMA is a partnership of ESO (representing its member states), NSF (USA) and NINS (Japan), together with NRC (Canada) and NSC and ASIAA (Taiwan), in cooperation with the Republic of Chile. The Joint ALMA Observatory is operated by ESO, AUI/NRAO and NAOJ. S.P.L. acknowledges support from the National Science Council of Taiwan with Grants NSC 98-2112-M-007- 007-MY3 and NSC 101-2119-M-007-004. Astrochemistry in Leiden is supported by the Netherlands Research School for Astronomy (NOVA), by a Spinoza grant and grant 614.001.008 from the Netherlands Organisation for Scientific Research (NWO), and by the European Community’s Seventh Framework Programme FP7/2007-2013 under grant agreement 238258 (LASSIE).
\end{acknowledgements}

\bibliographystyle{aa}
\bibliography{vla1623_disk_paper.bib}

\begin{appendix}
\section{VLA1623W}
\label{secW}
\begin{figure}
\centering
\includegraphics[angle=-90,scale=0.4]{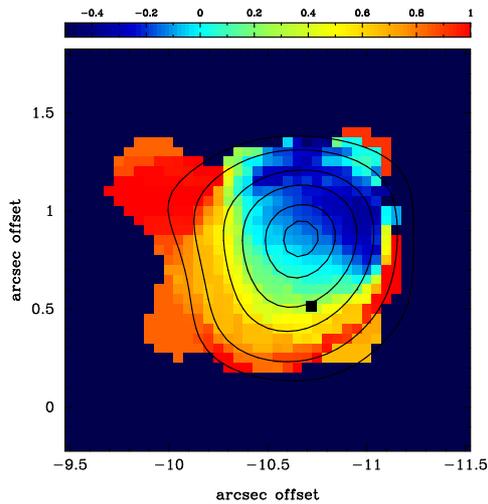}
\caption{Integrated intensity (Moment 0, contours) and velocity (Moment 1, halftone) maps of C$^{18}$O (2-1) detection towards VLA1623W. The position of VLA1623W is marked with a black square. Contours are in steps of 3$\sigma$, 5$\sigma$, 8$\sigma, $10$\sigma$, 12$\sigma$ and 13$\sigma$ with $\sigma$ = 13 mJy beam$^{-1}$.}
\label{figc18ow}
\end{figure}

In our ALMA observations we detect VLA1623W in continuum (see Section~\ref{seccont} and Figure~\ref{figcont}) and in C$^{18}$O line emission (Figure~\ref{figc18ow}), which was previously undetected with SMA observations. The emission is weak, peaking at 5$\sigma$ in the channel maps and at 13$\sigma$ in the intensity integrated map and has a velocity range of about -0.5 to 1 km s$^{-1}$. Noteworthy is VLA1623W's apparent systemic velocity $v_{\rm lsr}$ between 0 and 1 km s$^{-1}$, which differs from VLA1623A's systemic velocity by 3 to 4 km s$^{-1}$. VLA1623W's $v_{\rm lsr}$ is difficult to determine with certainty since the emission appears to be greatly affected by foreground material (see Section~\ref{secc18o}) and the low S/N of the detection. Finally, no C$^{18}$O emission was detected towards VLA1623B (Figure~\ref{figcont}, top left inset), consistent with previous CO depletion findings \citep{murillo2013}.

The velocity gradient of VLA1623W's C$^{18}$O emission appears to be consistent with a rotationally supported disk structure, and given that VLA1623W is classified as a Class I source \citep{murillo2013}, this is very likely. However, since the emission is very close to the edge of the field of view (Figure~\ref{figcont}) and is affected by foreground absorption, we are limited in carrying out kinematical analysis of the emission.

Finally the large difference in systemic velocity from VLA1623A and B may suggest one of two scenarios. First, that VLA1623W is not part of the system, but is instead a foreground or nearby source, that due to projection effects seems to be part of the system. However, \cite{dzib2013} rule out the possibility of VLA1623W (or VLA1623B) being a foreground or background object due to its proper motion. Second, that due to three body interaction, VLA1623W was ejected from the closer binary of VLA1623A and B \citep{reipurth2000}. This possibility is very likely given that VLA1623B may be a very recent formation, which would have caused the binary, now triple, to become unstable and eject one of the components. This ejection would cause VLA1623W to lose some of its envelope mass and appear more evolved, since one of the classification criteria is the envelope to central star mass ratio. The loss of mass would also make it more visible in the infrared as well as affect its evolution.

\Online

\section{Comparison of thin disk models and observations}
\label{seccomp}
In this appendix, we present the channel map comparison for each model and observations, as well as the residual obtained from subtracting the model from the observations. The figures presented below are a supplement to Figure~\ref{figmod} in order to gauge the best fitting model of the observed emission. All figures below show the blueshifted emission in the velocity range of 2.3 to 3 km s$^{-1}$, while the redshifted emission is in the velocity range of 4.8 to 6.5 km s$^{-1}$.

\begin{figure*}
\centering
\includegraphics[angle=-90,width=\linewidth]{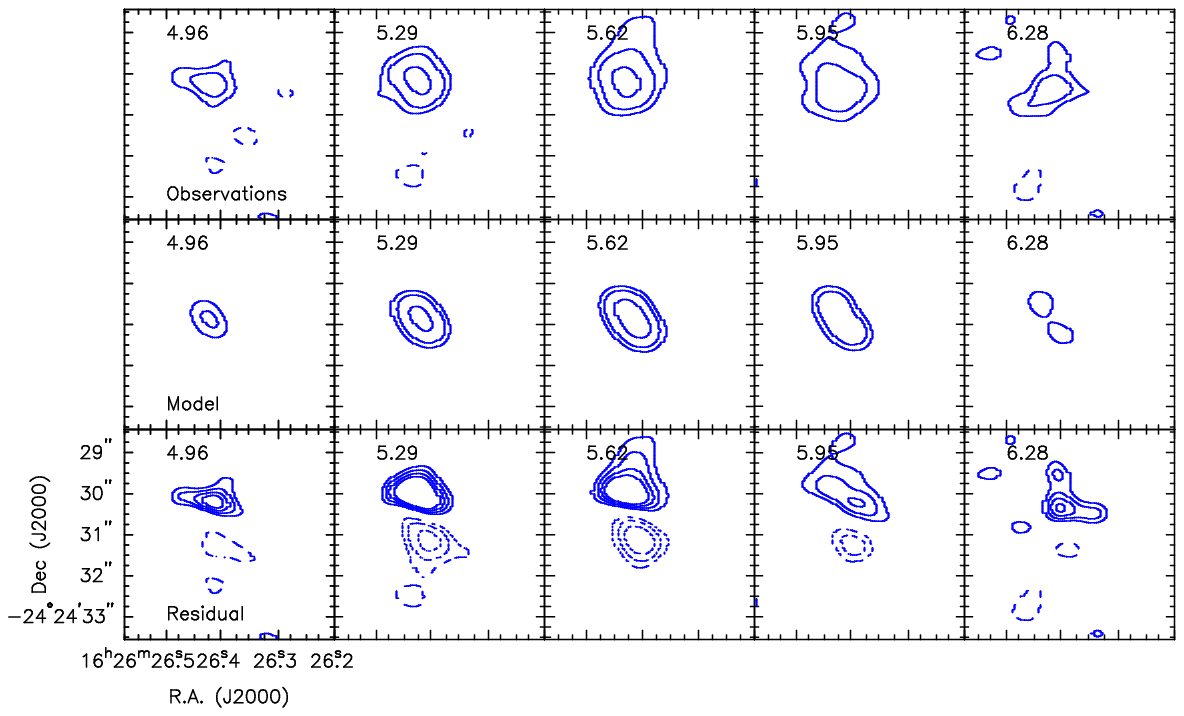} \\
\includegraphics[angle=-90,width=\linewidth]{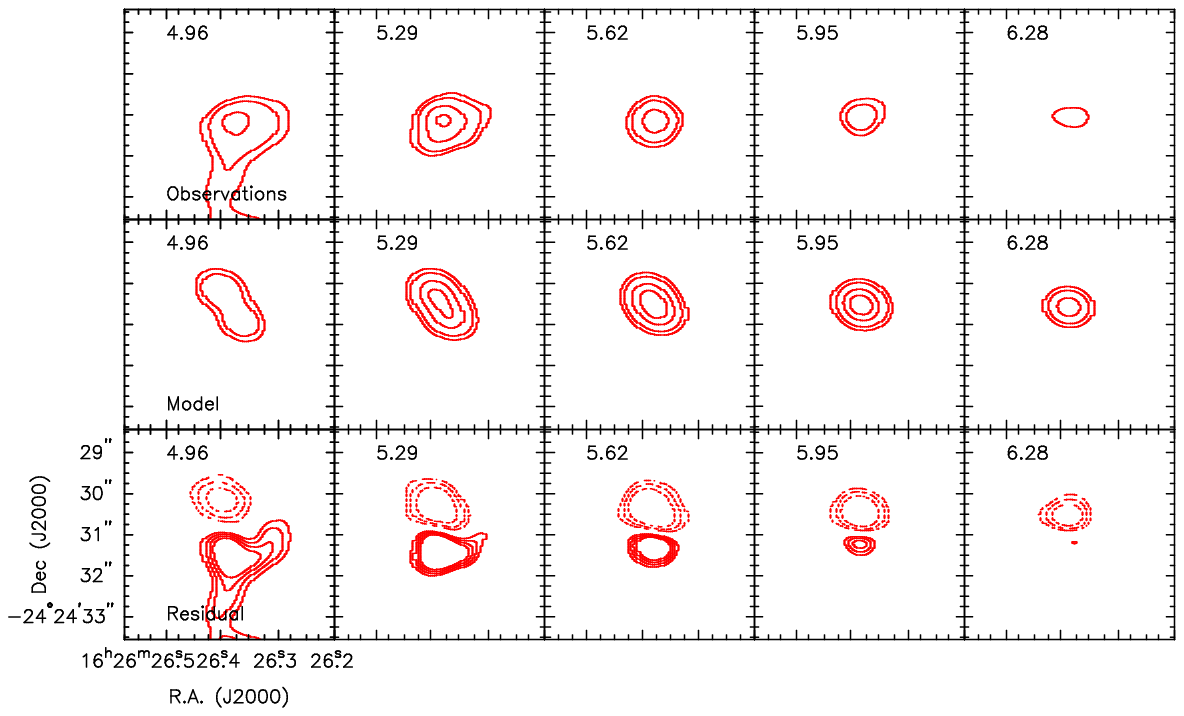}
\caption{Channel maps of C$^{18}$O observations (top), free falling disk model shown in Figure~\ref{figmod} (middle) and residual (bottom panel). The blue contours show the blueshifted emission and the red contours the redshifted emission. Number labels on the top of each panel indicate the velocity of that channel. Channels are binned to a velocity resolution of 0.16 km s$^{-1}$ for the blueshifted emission, and 0.33 km s$^{-1}$ for the redshifted emission for better display. Contours are in steps of 3$\sigma$, 5$\sigma$, 10$\sigma$, 15$\sigma$, 20$\sigma$ and 25$\sigma$ for the observations and model channel maps, and -8$\sigma$, -5$\sigma$, -3$\sigma$, 3$\sigma$, 4$\sigma$, 5$\sigma$ and 6$\sigma$ for the residual channel map, where $\sigma$ = 19 mJy beam$^{-1}$.}
\label{figresff}
\end{figure*}

\begin{figure*}
\centering
\includegraphics[angle=-90,width=\linewidth]{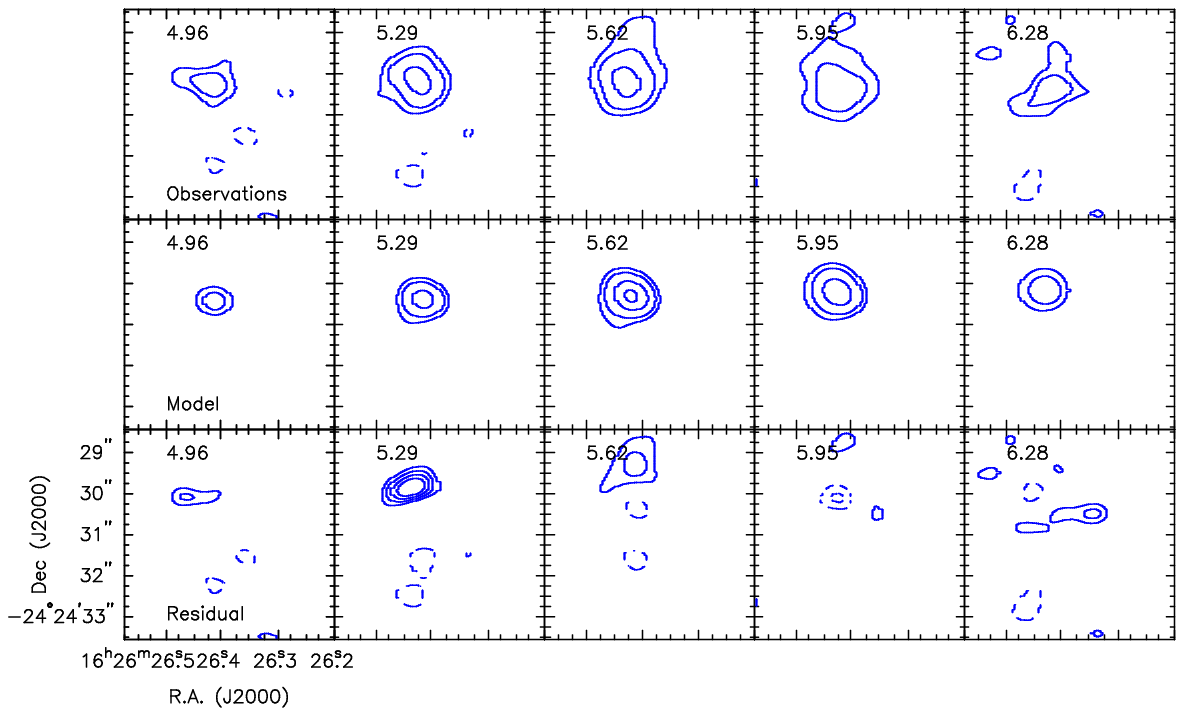} \\
\includegraphics[angle=-90,width=\linewidth]{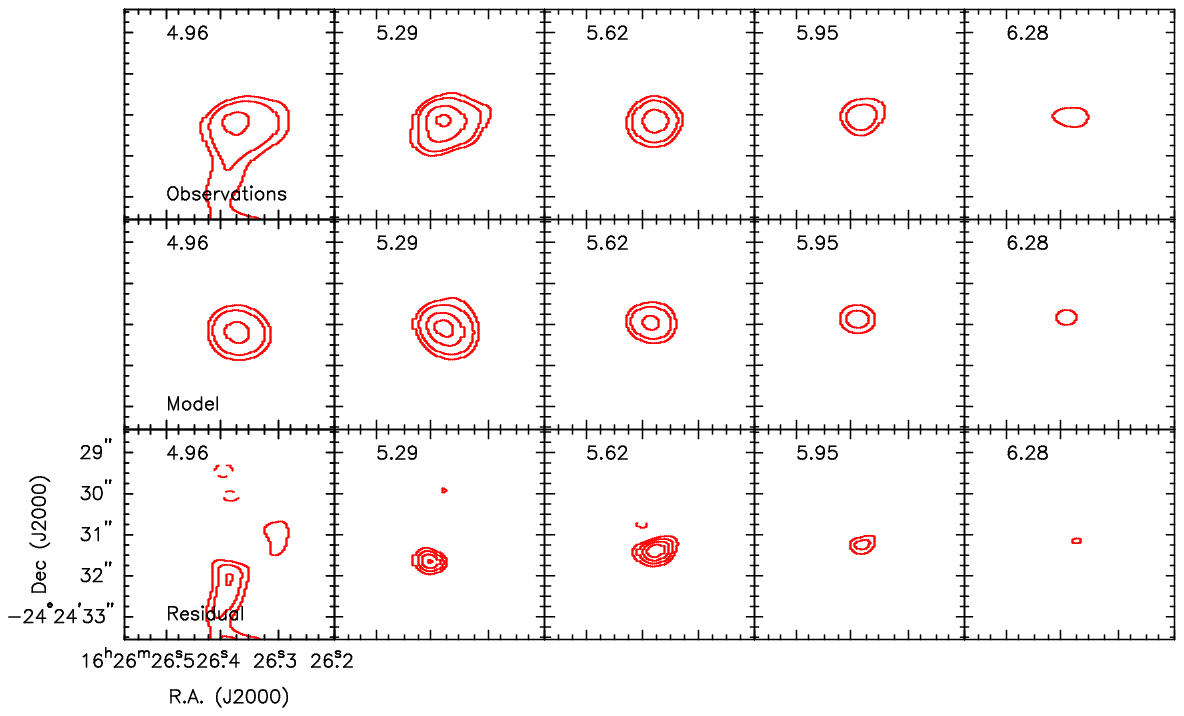}
\caption{Same as Figure~\ref{figresff} but for the free fall plus Keplerian disk model shown in Figure~\ref{figmod}.}
\label{figresffk}
\end{figure*}

\begin{figure*}
\centering
\includegraphics[angle=-90,width=\linewidth]{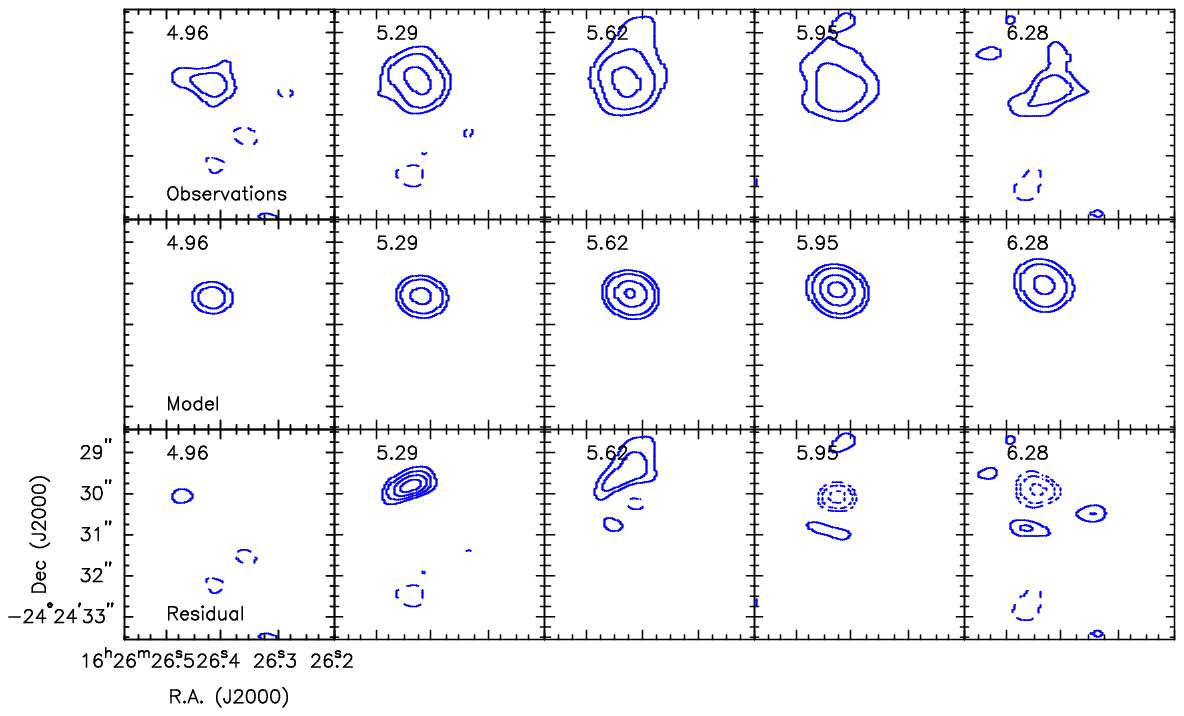} \\
\includegraphics[angle=-90,width=\linewidth]{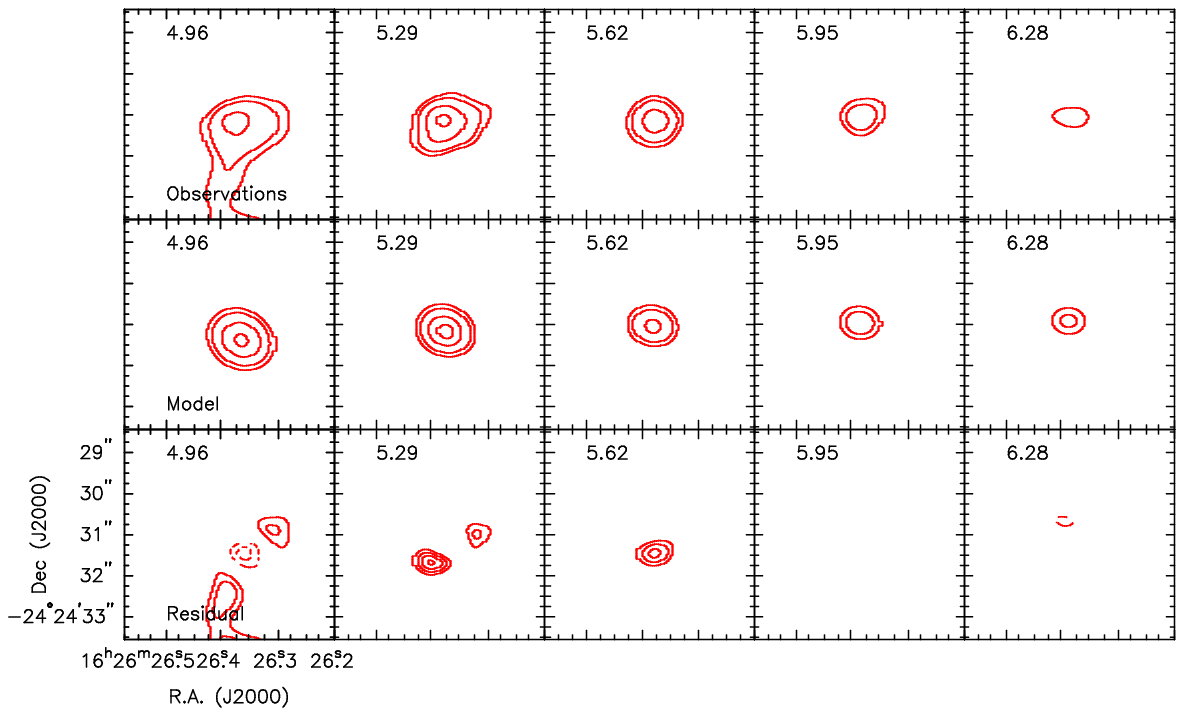}
\caption{Same as Figure~\ref{figresff} but for the Conserved Angular Momentum disk model shown in Figure~\ref{figmod}. Note the large negative residuals in the blueshifted emission.}
\label{figrescam}
\end{figure*}

\begin{figure*}
\centering
\includegraphics[angle=-90,width=\linewidth]{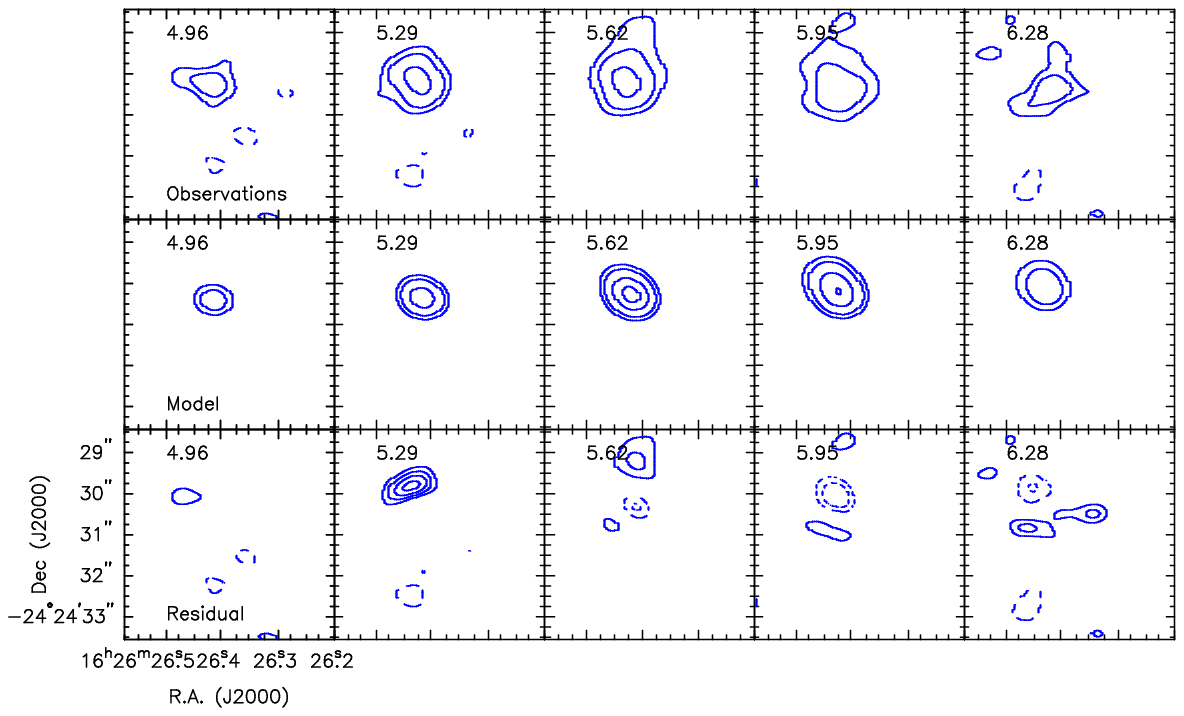} \\
\includegraphics[angle=-90,width=\linewidth]{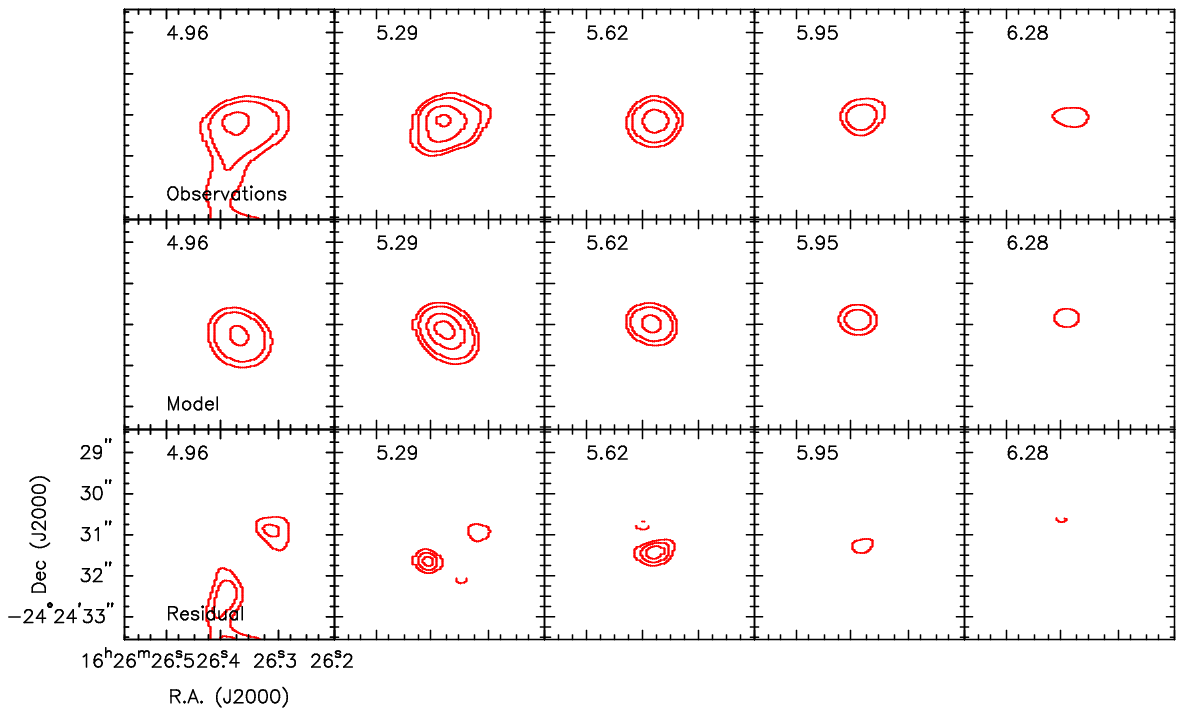}
\caption{Same as Figure~\ref{figresff} but for the Infall plus Keplerian disk model shown in Figure~\ref{figmod}.}
\label{figresink}
\end{figure*}

\begin{figure*}
\centering
\includegraphics[angle=-90,width=\linewidth]{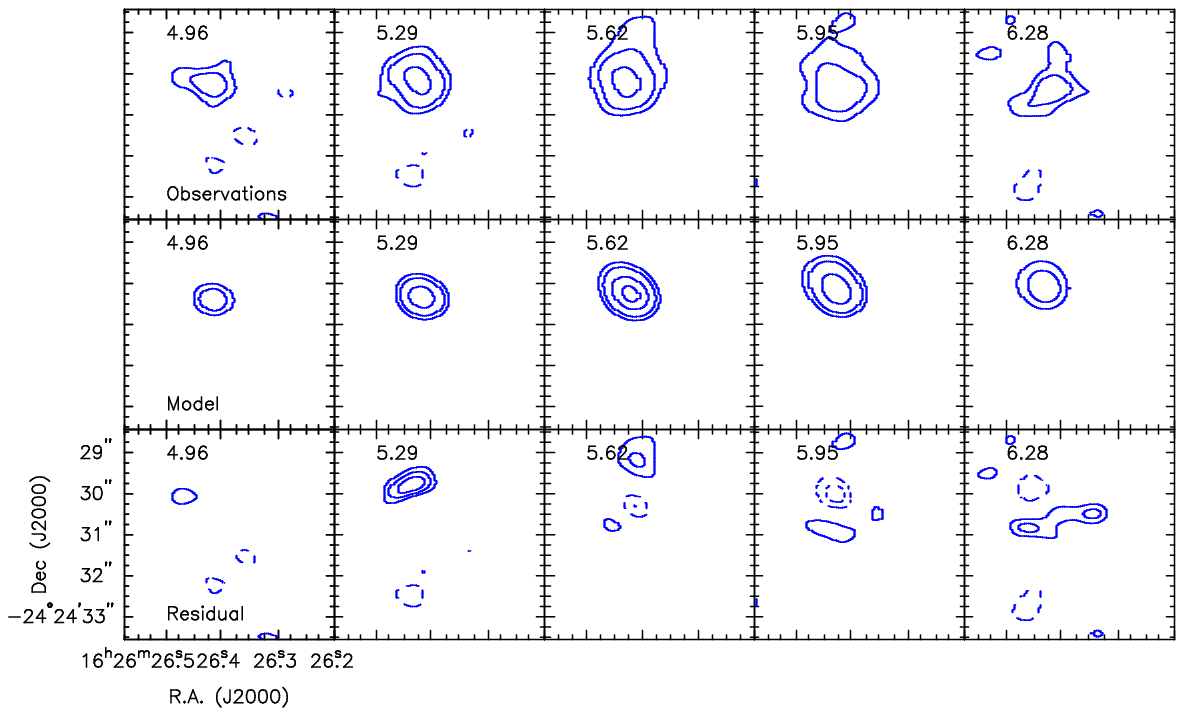} \\
\includegraphics[angle=-90,width=\linewidth]{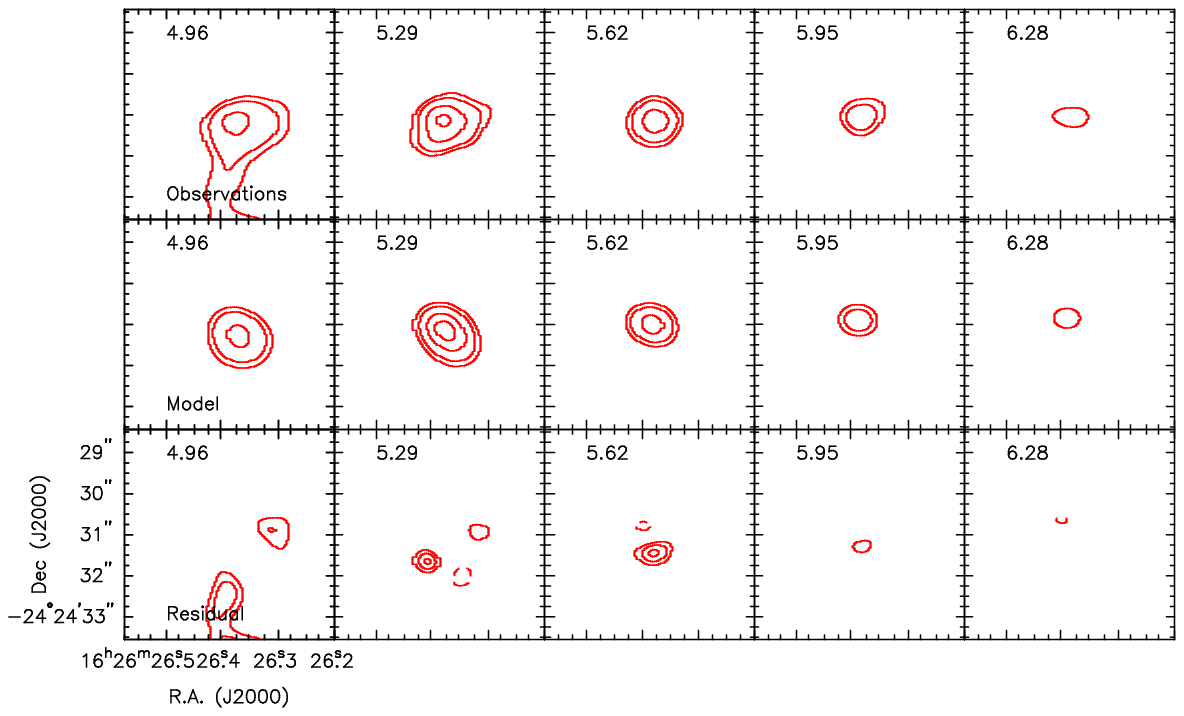}
\caption{Same as Figure~\ref{figresff} but for the best fit pure Keplerian disk model shown in Figure~\ref{figmod}.}
\label{figreskep}
\end{figure*}

\end{appendix}

\end{document}